\documentclass[aps,twocolumn,pra]{revtex4-1}

\usepackage{newlfont}
\usepackage{color}           
\usepackage{soul}            
\usepackage{amssymb}
\usepackage{amsfonts}
\usepackage{amsmath}
\usepackage{wasysym}
\usepackage{graphicx}
\usepackage{epsfig}
\usepackage{amsthm}
\usepackage{bm}
\usepackage{bbold}
\usepackage{epstopdf}

\usepackage[colorlinks=true, citecolor=blue, urlcolor=blue ]{hyperref}

\begin{document}

\definecolor{red}{rgb}{1,0,0}
\definecolor{orange}{rgb}{1,0.5,0}
\definecolor{blue}{rgb}{0,0,1}

\title{Distributed quantum dense coding with two receivers in noisy environments }

\author{Tamoghna Das, R. Prabhu, Aditi Sen(De), and Ujjwal Sen}

\affiliation{Harish-Chandra Research Institute, Chhatnag Road, Jhunsi, Allahabad 211 019, India}

\begin{abstract}
We investigate the effect of noisy channels in a classical information transfer through a multipartite state which acts as a substrate for the distributed quantum dense coding protocol between several senders and \emph{two} receivers. The situation is qualitatively different from the case with one or more senders and a single receiver. We obtain an upper bound on the multipartite capacity which is tightened in case of the covariant noisy channel. We also establish a relation between the genuine multipartite entanglement of the shared state and the capacity of distributed dense coding using that state,  both in the noiseless and the noisy scenarios. Specifically, we find that in the case of multiple senders and two receivers, the corresponding generalized Greenberger-Horne-Zeilinger states possess higher dense coding capacities as compared to a significant fraction of pure states having the same multipartite entanglement.
\end{abstract}

\maketitle

\section{Introduction}
Quantum entanglement is one of the essential ingredients in quantum information processing tasks which include  superdense coding  \cite{BennetDC}, teleportation \cite{TeleRef}, quantum error correction \cite{Quantum_error}, quantum secret sharing \cite{QCryp, qsecretsharing}, and one way quantum computation \cite{onewayQC}. It was shown that such protocols can provide advantage over the corresponding  classical protocols \cite{amaderreview}. Moreover, classical information as well as quantum state transfer via quantum channels have been successfully realized in the laboratory over reasonably large distances \cite{DCexp, teleexp}. 

Any communication protocol involves three major steps -- (1) encoding of the information in a physical system, (2) sending the physical system through a physical channel, and (3) decoding the information. In this paper, we are interested in the communication scheme which deals with the transfer of classical information encoded in a quantum state shared between distant parties, and is known as quantum dense coding (DC) \cite{BennetDC, dcgeneral}. Capacity of the dense coding protocol have been evaluated in several scenarios involving a single receiver. These include the cases of a single sender as well as multiple senders and in both noiseless and noisy scenarios \cite{BennetDC, dcgeneral, DistriDC, Dense_multi,Shadmansob}. An important tool here is the Holevo bound on the accessible information for ensembles of quantum states \cite{HolevoQuantity1,HolevoQuantity2}. The situations when there are multiple senders and/or multiple receivers are involved, have been termed as distributed quantum dense coding \cite{DistriDC,Dense_multi}. The capacity in the noiseless case for two receivers has been estimated in  \cite{DistriDC,Dense_multi}, where the Holevo-like upper bound on locally accessible information for ensembles of quantum states of bipartite systems was used \cite{AccLOCC}.

In this paper, we estimate the capacity of distributed quantum dense 
coding for two receivers in the noisy case. The receivers are allowed to
perform local (quantum) operations and classical communication (LOCC), and
we term the communication protocol as the LOCC-DC protocol and the
corresponding capacity as the LOCC-DC capacity. We begin by finding an
upper bound on the capacity for arbitrary noisy channels between the
senders and the receivers. A tighter bound in closed form is obtained for
the case of covariant channels \cite{covariancenoise}. When
the shared state is a Greenberger-Horne-Zeilinger (GHZ) state \cite{GHZ} and
when the noisy channels are among the amplitude damping, phase damping, or
the Pauli channels, the upper bounds on the LOCC-DC capacities are explicitly evaluated. Furthermore, we relate the LOCC-DC capacity with the multiparty entanglement in the shared state, in both noiseless and noisy
cases. We had recently observed in the case of several senders and a
single receiver that noise in the channel inverts relative capability of
information transfer in dense coding between generic multiparty pure
quantum states and the corresponding generalized GHZ (gGHZ) states \cite{Noiseinvert}.
Here we find that such inversion does not take place in the case of two
receivers (and several senders): The gGHZ state provides better
classical information-carrying capacity for both noiseless and noisy cases in
comparison to a significantly high fraction of pure states in the
corresponding Hilbert space.

The paper is organized as follows. In Sec. \ref{Sec:Densecoding}, we discuss the multiparty DC capacity for more than one receiver, with the decoding operations being restricted to LOCC. In the case of multiple senders and two receivers, we establish an upper bound on the DC capacity for noisy quantum channels. A tighter upper bound on the LOCC-DC capacity in presence of covariant noise is obtained in Sec.   \ref{subsubsec:Covariant}.  In Sec. \ref{Sec:exmaple_noise}, we evaluate closed forms of LOCC-DC capacity for some specific noisy quantum channels, when a four-qubit GHZ state is shared.  In Sec. \ref{Sec:GGMDef}, we briefly introduce the generalized geometric measure (GGM), a genuine multiparty entanglement measure. We establish connections between the entanglement measure with the upper bound on information transfer  in Sec. \ref{Sec:Relationship_GGM_Capacity}. 
Finally, we present a conclusion in Sec \ref{Sec:Conclusion}.

 \section{Quantum dense coding for more than one Receiver}
\label{Sec:Densecoding}

We consider the quantum  dense coding protocol with an arbitrary number of senders and two receivers. Let an $(N+2)$-party quantum state, $\rho^{S_1S_2\ldots S_NR_1R_2}$, be shared between $N$ senders, $S_1, S_2,\ldots, S_N $, and two receivers, $R_1$ and $R_2$. And among them, some of the senders send their encoded quantum state to the first receiver while the rest will send to the second receiver, through noiseless or noisy quantum channels.

\begin{figure}
\includegraphics[width=1.15\columnwidth,keepaspectratio,angle=0]{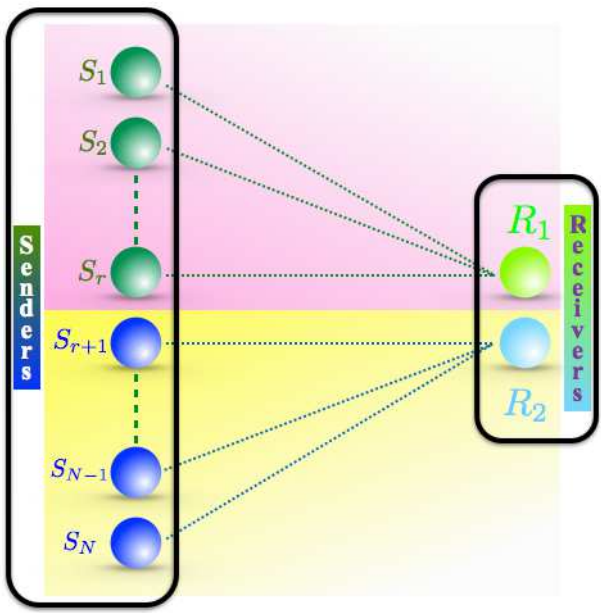}
\caption{(Color online.) A schematic diagram of the DC protocol considered.  An $(N+2)$-party quantum state, $\rho^{S_1S_2\ldots S_NR_1R_2}$, is shared between $N$ senders, $S_1, S_2,\ldots, S_N $, and two receivers, $R_1$ and $R_2$. We assume that  after unitary encoding, the senders, $S_1,S_2,\ldots, S_r$, send their part to the receiver $R_1$ while the rest send their parts to the receiver, $R_2$. }
\label{schematic_2receivers}
\end{figure}

The amount of classical information that the senders can send to the receivers depends on four factors -- $(1)$ encoding procedures used by the senders, \((2)\) the probability of the sampling rate of different encodings, $(3)$ properties of channels by which the encoded states have to be sent, and $(4)$ the measurement strategies used by the receivers to decode the message. Let us first concentrate on the case when the decoding procedures which the receivers are allowed to make are global operations. The capacity of dense coding, in this case, reduces to optimization of the Holevo quantity over unitary encodings (for different encodings, see \cite{HoroPiani}) and probabilities. The multiparty DC capacity for an arbitrary multiparty state $\rho^{S_1\ldots S_NR_{1}R_2}$, with N senders, $S_1, S_2, \ldots,$ and $ S_N$, and two receivers, \(R_1\) and \(R_2\), who are in this case together and denoted by $R= R_1 R_2$, is given by \cite{dcgeneral, DistriDC, Dense_multi}
\begin{equation*}
C^G = \log d_{S_1S_2\ldots S_N} + S(\rho^R) -S(\rho^{S_1S_2\ldots S_NR}),
\end{equation*} 
where $d_{S_1S_2\ldots S_N}$ is the dimension of the Hilbert space of all the senders, and $S(\sigma) = - \text{tr}(\sigma\log\sigma)$  is the von Neumann entropy of the density matrix, $\sigma$.  In this paper all logarithms will be with base 2, and there by all capacities will be measured in bits.
Such a situation can arise, e.g., if  $R_1$ teleports his \cite{Female}  quantum systems to $R_2$ after obtaining the post-encoded systems from the senders.

The case when the receivers are at distant locations and when teleportation or global operations are not allowed, has a further two possibilities. $(i)$ When the receivers are not allowed to communicate  between themselves, the corresponding DC capacity is additive with respect to the receivers, and is known as the LO-DC  protocol \cite{DistriDC,Dense_multi}. 
  $(ii)$ When the receivers are allowed to perform LOCC for decoding, the protocol is called LOCC-DC. It is the second case which is considered in this paper, and we will now describe it in some detail. Consider again a multiparty state, $\rho^{S_1 \ldots S_N R_1R_2}$, shared between N senders and two receivers, $R_1$ and $R_2$. To send the classical information, $\{i\}$, which occurs with probability $p_{\{i\}} = p_{i_1}\ldots p_{i_r}p_{i_{r+1}}\ldots p_{i_N}$, some of the senders, say, $S_1,S_2,\ldots,S_r$, perform  either local  or global unitary operations, denoted by $U_{i_1\ldots i_r}^{S_1\ldots S_r}$ with probabilities $p_{i_1\ldots i_r}$, on their parts of the shared state and send it to the receiver $R_1$.
The rest of the senders, $S_{r+1},S_{r+2},\ldots, S_N$, also perform unitary operations, $U_{i_{r+1}\ldots i_N}^{S_{r+1}\ldots S_N}$, with probabilities $p_{i_{r+1}\ldots i_N}$ on their parts and send it to $R_2$ (see  Fig. \ref{schematic_2receivers}). Finally, the receivers, $R_1$ and $R_2$ 
possess an ensemble of state $\{p_{\{i\}}, \rho^{S_1\ldots S_NR_1R_2}_{\{i\}}\}$, where $p_{\{i\}} = p_{i_1\ldots i_r}\times p_{i_{r+1}\ldots i_N}$, and $\rho^{S_1\ldots S_NR_1R_2}_{\{i\}} = U_{i_1\ldots i_r}^{S_1\ldots S_r}\otimes U_{i_{r+1}\ldots i_N}^{S_{r+1}\ldots S_N}\otimes \text{I}_{R_1} \otimes \text{I}_{R_2}\rho^{S_1\ldots S_NR_1R_2} U_{i_1\ldots i_r}^{S_1\ldots S_r \dagger}\otimes U_{i_{r+1}\ldots i_N}^{S_{r+1}\ldots S_N \dagger}\otimes \text{I}_{R_1} \otimes \text{I}_{R_2}$, with $\text{I}_{R_1}$ and $\text{I}_{R_2}$ being the identity operators in the receiver Hilbert spaces. The receivers, $R_1$ and $R_2$, now apply an LOCC protocol in the $S_1\ldots S_rR_1:S_{r+1}\ldots S_NR_2$ bipartition to decode the information that the senders have sent.

The LOCC-DC protocol can be considered for the noiseless channel \cite{DistriDC,Dense_multi}, or when the channels from the senders and the receivers are noisy.  We first deal with the general noisy channel and then consider the covariant channel.
 
 \subsection{Capacity of Dense Coding for Many Senders and Two Receivers -- Noisy Channels }\label{Subsec:Densecodingcapacity_noisy}
 In this section, our aim is to estimate the capacity when multiple senders send their encoded parts of the shared quantum state to the two receivers by using a general noisy quantum channel. In a realistic situation, the transmission channel can not be kept completely  isolated from the environment, and hence noise almost certainly acts on the  encoded parts of the senders' side while sending their parts through the shared channels.

   Mathematically, noise in the transmission channel is a completely positive trace preserving map (CPTP), $\Lambda$, acting on the state space of the senders' part of the transmitted state. Therefore, the receivers, $R_1$ and $R_2$, after the transmission,  possess the distorted ensemble, $\{p_{\{i\}}, \Lambda_{S_1\ldots S_N}(\rho_{\{i\}}^{S_1\ldots S_NR_1R_2})\}$, in the $S_1\ldots S_rR_1:S_{r+1}\ldots S_NR_2$ bipartition, where $\Lambda_{S_1\ldots S_N}(\rho_{\{i\}}^{S_1\ldots S_NR_1R_2}) = \Lambda_{S_1\ldots S_N}(U_{i_1\ldots i_r}^{S_1\ldots S_r}\otimes U_{i_{r+1}\ldots i_N}^{S_{r+1}\ldots S_N}\otimes \text{I}_{R_1} \otimes \text{I}_{R_2}\rho^{S_1\ldots S_NR_1R_2} U_{i_1\ldots i_r}^{S_1\ldots S_r \dagger}\otimes U_{i_{r+1}\ldots i_N}^{S_{r+1}\ldots S_N \dagger}\otimes \text{I}_{R_1} \otimes \text{I}_{R_2})$. 
 To estimate the capacity,  the $(N+2)$-party quantum state, $\rho^{S_1\ldots S_NR_1R_2}$, can be expanded as 
 

%

\begin{eqnarray}\label{Eq:rhoexpansion}
 \rho^{S_1\ldots S_NR_1R_2} = \sum_{\{i,j\}} &&\lambda_{\{i,j\}}|i_1\rangle\langle j_1|^{S_1\ldots S_N}\otimes |i_2\rangle\langle j_2|^{R_1} \nonumber \\ &&\hspace{7em}\otimes |i_3\rangle \langle j_3|^{R_2} ,
\end{eqnarray}
where $\{|i_1\rangle\}_{i_1 = 0}^{d_{S_1\ldots S_N}-1}$, $\{|i_2\rangle\}_{i_2 = 0}^{d_{R_1}-1}$, and $\{|i_3\rangle\}_{i_3 = 0}^{d_{R_2}-1}$  are respectively bases in the Hilbert space $ {\cal H}^{S_1\ldots S_N}$, of all the senders, and ${\cal H}^{R_1}\, ({\cal H}^{R_2})$ of the receiver $R_1\, (R_2)$.


After the action of the CPTP map, $\Lambda$, on the encoded state, we get
\begin{eqnarray}\label{Eq:CPTPexpansion}
 &&\Lambda_{S_1\ldots S_N}(\rho^{S_1\ldots S_NR_1R_2}_{\{i\}}) = \sum_{\{i,j\}} \lambda_{\{i,j\}}\nonumber \\ &&\Lambda_{S_1\ldots S_N}(U_{i_1\ldots i_r}^{S_1\ldots S_r}\otimes U_{i_{r+1}\ldots i_N}^{S_{r+1}\ldots S_N}|i_1\rangle\langle j_1|^{S_1\ldots S_N}U_{i_1\ldots i_r}^{S_1\ldots S_r \dagger}\nonumber \\ && \hspace{4em} \otimes U_{i_{r+1}\ldots i_N}^{S_{r+1}\ldots S_N \dagger})  \otimes |i_2\rangle\langle j_2|^{R_1}\otimes |i_3\rangle \langle j_3|^{R_2},
\end{eqnarray}
where $\Lambda_{S_1\ldots S_N}$ is collectively or individually acting only on the senders' subsystems.

The amount of classical information that can be extracted from the ensemble, $\{p_{\{i\}}, \Lambda_{S_1\ldots S_N}(\rho_{\{i\}}^{S_1\ldots S_NR_1R_2})\}$, by LOCC, is given by \cite{AccLOCC}  
  \begin{equation}\label{Eq:Acceinequality}
   \text{I}^{\text{LOCC}}_{acc} \leq S(\xi^1) + S(\xi^2) - \max_{x\in1,2}\sum_{\{i\}}p_{\{i\}}S(\xi_{\{i\}}^x),
  \end{equation}
where $\xi^1_{\{i\}} = \text{tr}_{S_{r+1}\ldots S_NR_2}\Lambda(\rho_{\{i\}}^{S_1\ldots S_NR_1R_2})$, $\xi^2_{\{i\}} = \text{tr}_{S_{1}\ldots S_rR_1}\Lambda(\rho_{\{i\}}^{S_1\ldots S_NR_1R_2})$ and $\xi^{1,2} = \sum_{\{i\}} p_{\{i\}}\xi^{1,2}_{\{i\}}$.
The Holevo bound \cite{HolevoQuantity1} on accessible information is asymptotically achievable \cite{HolevoQuantity2}. However, for two receivers \cite{AccLOCC}, such asymptotic achievability has not yet been proven. Therefore, unlike the cases when a single receiver is involved \cite{BennetDC,dcgeneral,DistriDC,Dense_multi,Shadmansob}, the LOCC-DC capacity can only be estimated with an upper bound \cite{DistriDC,Dense_multi}.

To obtain the capacity of LOCC-DC in the noiseless scenario, one has to maximize the right-hand-side (R.H.S.) of inequality (\ref{Eq:Acceinequality}) over unitary encodings and probabilities with $\Lambda = I$
and we obtain \cite{DistriDC,Dense_multi}
\begin{eqnarray}\label{Eq:Capacitydef}
  C^{\text {LOCC}} \leq \log d_{S_1\ldots S_N} + S(\rho^{R_1})+  S(\rho^{R_2})   \nonumber \\  - \max_{x=1,2}S(\rho^x) \equiv {\cal B}^{\text{LOCC}},
\end{eqnarray}
where $\rho^{R_i} = \text{tr}_{S_1\ldots S_NR_j}\rho^{S_1\ldots S_NR_1R_2}$ with $i,j = 1,2$, $i\neq j$, and $\rho^1 = \text{tr}_{{S_{r+1}\ldots S_NR_{2}}}\rho^{S_1\ldots S_NR_1R_2} $, $\rho^2 = \text{tr}_{{S_1\ldots S_rR_{1}}}\rho^{S_1\ldots S_NR_1R_2} $.

 
Like in the noiseless case, to obtain the capacity of LOCC-DC in a noisy scenario, one has to maximize the R.H.S. of (\ref{Eq:Acceinequality}) over unitaries and probabilities. The ensemble in the noisy scenario, involve the CPTP map $\Lambda$.
 \begin{eqnarray}\label{Eq:maximize}
 C_{noisy}^{\text{LOCC}} \leq \chi^{\text{LOCC}}_{noisy} &=& \max[S(\xi^1) + S(\xi^2) \nonumber \\ &&\hspace{5mm}- \max_{x\in1,2}\sum_{\{i\}}p_{\{i\}}S(\xi_{\{i\}}^x)]
 \end{eqnarray}

\noindent If we apply the subadditivity of von Neumann entropy in the $S_1\ldots S_r:R_1$ and $S_{r+1}\ldots S_N:R_2$ bipartitions for the first two terms, we have
\begin{equation}\label{Eq:xi_inequality}
 S(\xi^k)\leq S(\xi^{k'}) + S(\xi^{k''}) \leq \log d_{\bar{R}_k} + S(\rho^{R_k}), \hspace{3mm} k=1,2
\end{equation}
where $\xi^{k'} = \text{tr}_{R_k}\xi^k$ and $\xi^{k''} = \text{tr}_{\bar{R}_k}\xi^k = \rho^{R_k}$, with $\bar{R}_1 = S_1 \ldots S_r $,  $\bar{R}_2 = S_{r+1} \ldots S_N$. 
The second inequality is due to the fact that the maximum von Neumann entropy of a $d$-dimensional quantum state is $\log d$.


To deal with  the third term in the R.H.S. of (\ref{Eq:maximize}), let us assume that $U_{min}^{S_1\ldots S_r}$ and $U_{min}^{S_{r+1}\ldots S_N}$ are two unitary operators acting on  subsystems $S_1\ldots S_r$ and $S_{r+1}\ldots S_N$  of $\rho^{S_1\ldots S_NR_1R_2}$ respectively. Let us suppose that after passing through the noisy transmission channel $\Lambda_{S_1\ldots S_N}$, those unitaries give the minimum von Neumann entropy among all the von Neumann entropies of $\xi_{\{i\}}^k, k=1,2$, of the ensemble. 
Consider $\tilde{\rho}^{S_1\ldots S_NR_1R_2} = U_{min}^{S_1\ldots S_r}\otimes U_{min}^{S_{r+1}\ldots S_N}\otimes \text{I}^{R_1}\otimes \text{I}^{R_2} \rho^{S_1\ldots S_NR_1R_2} U_{min}^{S_1\ldots S_r \dagger}\otimes U_{min}^{S_{r+1}\ldots S_N \dagger}\otimes \text{I}^{R_1}\otimes \text{I}^{R_2}$, and the corresponding reduced density matrices
\begin{eqnarray}\label{Eq:zetaOne}
 \zeta^1 = \text{tr}_{S_{r+1}\ldots S_NR_2}\Lambda_{S_1\ldots S_NR_1R_2}(\tilde{\rho}^{S_1\ldots S_NR_1R_2}),
 \end{eqnarray}
 \begin{eqnarray}\label{Eq:zetaTwo}
 \zeta^2 = \text{tr}_{S_{1}\ldots S_rR_1}\Lambda_{S_1\ldots S_NR_1R_2}(\tilde{\rho}^{S_1\ldots S_NR_1R_2}).
\end{eqnarray}
Since entropy is concave,
one should expect that the set, $\{S(\xi^x_{\{i\}})\}$, of real numbers, which depend on the unitary operators $U_{i_1\ldots i_r}^{S_1\ldots S_r}$ or $U_{i_{r+1}\ldots i_N}^{S_{r+1}\ldots S_N}$  must have a minimum value, 
denoted by $S(\zeta^x)$, which can be achieved by the unitary operators $U_{min}^{S_1\ldots S_r}$ and $U_{min}^{S_{r+1}\ldots S_N}$.
Hence we have
\begin{eqnarray}
S(\xi^x_{\{i\}}) &\geq& S(\zeta^x)\,\,\,\,\,\forall \,\,\,i 
\end{eqnarray}
which implies
\begin{eqnarray}
 \sum_{\{i\}}p_{\{i\}}S(\xi^x_{\{i\}}) &\geq& \sum_{\{i\}}p_{\{i\}} S(\zeta^x) = S(\zeta^x).
\end{eqnarray}

One should note here that  $U_{min}^{S_1\ldots S_r}$ and $U_{min}^{S_{r+1}\ldots S_N}$ independently minimize $S(\zeta^1)$ and $S(\zeta^2)$, respectively. For example, to minimize the von Neumann entropy, of $\xi_{\{i\}}^1$,  we already traced out the other partition of $\rho^{S_1\ldots S_NR_1R_2}$ and $U_{min}^{S_{r+1}\ldots S_N}$ and hence the minimization procedure in $\sum_{i}p_{\{i\}}\xi_{\{i\}}^1$ depends only on $U_{min}^{S_1\ldots S_r}$.   Similar argument holds for $\sum_{i}p_{\{i\}}\xi_{\{i\}}^2$ also. Thus we have the following theorem.


\noindent{\bf Theorem 1:} \textit{For arbitrary noisy channels between multiple senders and the two receivers, the LOCC dense coding capacity, involving two receivers, is bounded above by the quantity}
\begin{eqnarray}\label{Eq:Theorem_statement}
 {\cal B}^{\text{LOCC}}_{noisy} \equiv \log d_{S_1\ldots S_N} + S(\rho^{R_1}) + S(\rho^{R_2}) \nonumber \\ - \max_{x\in1,2}S(\zeta^x).
\end{eqnarray}
Here $\zeta^1$ and $\zeta^2$ are  respectively given in Eqs. (\ref{Eq:zetaOne}) and (\ref{Eq:zetaTwo}).
The question remains whether there exists any noisy channel for which the upper bound can be made tighter than the one given in Eq. (\ref{Eq:Theorem_statement}). We will address the question below.

\subsubsection{Covariant Noisy Channel}\label{subsubsec:Covariant}
 We will now deal with a class of noisy channels called the covariant channels. For an arbitrary quantum state $\rho$ in $d$ dimensions, the CPTP map,  $\Lambda^C$, is said to be covariant, if one can find a  complete set of orthogonal unitary operators, $\{W_i\}_{i=0}^{d^2-1}$, acting on the state space of $\rho$, such that
 \begin{equation}\label{Eq:Covariant_def}
  \Lambda^C(W_i\rho W_i^{\dagger}) = W_i\Lambda^C(\rho)W_i^{\dagger}, \,\,\,\,\,\forall \,\,\, i,
 \end{equation}
$\{W_i\}$ satisfies the orthogonality condition, given by
\begin{equation}\label{Eq:UnitaryOrthogonalCondition}
 \frac{1}{d}\text{tr}(W_iW_j^{\dagger}) = \delta_{ij},
\end{equation}
and the completeness relation 
 \begin{equation}\label{Eq:UnitaryCompleteness}
 \frac{1}{d}\sum_iW_i\Xi W_i^{\dagger} = \text{I}_d \text{tr}\Xi,
 \end{equation}
where $\Xi$ is any operator in the same Hilbert space as $\rho$.
After encoding at the senders' side, we assume that the senders' part are sent through the noisy covariant channel, $\Lambda^C_{S_1 \ldots S_N}$. After passing through the channel, the resulting state is given by
 %
 %
%
\begin{eqnarray}\label{Eq:Coexpansion}
 &&\Lambda_{S_1\ldots S_N}^C(\rho^{S_1\ldots S_NR_1R_2}_{\{i\}}) = \sum_{\{i,j\}} \lambda_{\{i,j\}}\nonumber \\ &&\Lambda_{S_1\ldots S_N}^C(U_{i_1\ldots i_r}^{S_1\ldots S_r}\otimes U_{i_{r+1}\ldots i_N}^{S_{r+1}\ldots S_N}|i_1\rangle\langle j_1|^{S_1\ldots S_N}U_{i_1\ldots i_r}^{S_1\ldots S_r \dagger}\nonumber \\ && \hspace{4em} \otimes U_{i_{r+1}\ldots i_N}^{S_{r+1}\ldots S_N \dagger})  \otimes |i_2\rangle\langle j_2|^{R_1}\otimes |i_3\rangle \langle j_3|^{R_2},
\end{eqnarray}
where we use the form of an arbitrary $(N+2)$-party quantum state given in Eq. (\ref{Eq:rhoexpansion}), and, $\Lambda_{S_1\ldots S_N}^C$, is a covariant noise acting on the state space of $S_1\ldots S_N$,  satisfying Eq. (\ref{Eq:Covariant_def}), with the complete set of orthogonal unitary operators belonging to the linear operator space ${\cal L}({\cal H}^{S_1\ldots S_N})$. We are going to show that in this case, the maximization involved in the upper bound on the capacity depends on the fixed unitary operator and the Kraus operator of the channel $\Lambda_{S_1\ldots S_N}^C$.	

Let $\{V_j^{S_1\ldots S_r}\}_{j=0}^{d^2_{S_1 \ldots S_r} - 1}$, with probabilities $p_{j} = \frac{1}{d^2_{S_1\ldots S_r}}$, and $\{V_{j'}^{S_{r+1}\ldots S_N}\}_{j' = 0}^{d^2_{S_{r+1}\ldots S_N}-1} $ with probabilities $p_{j'} = \frac{1}{d^2_{S_{r+1}\ldots S_N}}$, be two complete sets of orthogonal unitary operators satisfying Eq. (\ref{Eq:UnitaryOrthogonalCondition}), respectively acting on the Hilbert spaces of the senders $S_1\ldots S_r$, and $S_{r+1}\ldots S_N$. Without loss of generality, we assume that the first bunch of senders send their encoded parts to the receiver $R_1$, while the rest sends to the receiver $R_2$. Let 


\begin{eqnarray}\label{Eq:ensemblestates}
 \rho_{j,j'}^{S_1\ldots S_NR_1R_2} &=& (V_j^{S_1\ldots S_r}\otimes V_{j'}^{S_{r+1}\ldots S_N }\otimes \text{I}_{R_1}\otimes \text{I}_{R_2}) \hspace{3em} \nonumber \\ && \hspace{-5em}\rho^{S_1\ldots S_NR_1R_2} (V_j^{S_1\ldots S_r \dagger}\otimes V_{j'}^{S_{r+1}\ldots S_N \dagger} \otimes \text{I}_{R_1}\otimes \text{I}_{R_2}).
\end{eqnarray}
One can always write $V_j^{S_1\ldots S_r} = W_j^{S_1\ldots S_r}U_1^{S_1\ldots S_r}$ and $V_{j'}^{S_{r+1}\ldots S_N } = W_{j'}^{S_{r+1}\ldots S_N }U_2^{S_{r+1}\ldots S_N }$, where $W_j^{S_1\ldots S_r} \otimes W_{j'}^{S_{r+1}\ldots S_N}$ acting on the senders state space, satisfying Eqs. (\ref{Eq:UnitaryOrthogonalCondition}) and (\ref{Eq:UnitaryCompleteness}), commutes with the covariant map, $\Lambda_{S_1 \ldots S_N}^C$, while   $U_1^{S_1 \ldots S_r}$ and $U_2^{S_{r+1} \ldots S_N}$ are fixed unitary operators. Therefore, after the encodings and passing through the covariant channel, the ensemble states of the DC protocol are


\begin{eqnarray}
 \Lambda_{S_1\ldots S_N}^C(\rho_{j,j'}^{S_1\ldots S_NR_1R_2}) =
 W_j^{S_1\ldots S_r}\otimes W_{j'}^{S_{r+1}\ldots S_N} \otimes \text{I}^{R_1} \nonumber \\
 \otimes \text{I}^{R_2} \Lambda_{S_1\ldots S_N}^C(U_1\otimes U_2\otimes \text{I}^{R_1} \otimes \text{I}^{R_2}\rho^{S_1\ldots S_NR_1R_2}\nonumber \\
 U_1^{\dagger}\otimes U_2^{\dagger}\otimes \text{I}^{R_1} \otimes \text{I}^{R_2}) W_j^{S_1\ldots S_r \dagger}\otimes W_{j'}^{S_{r+1}\ldots S_N \dagger} \otimes \text{I}^{R_1} \otimes \text{I}^{R_2}, \nonumber \\
\end{eqnarray}
where we have used the covariant condition, given in Eq. (\ref{Eq:Covariant_def}), on $\Lambda_{S_1\ldots S_N}^C$. Let us denote $\Lambda_{S_1\ldots S_N}^C (U_1 \otimes U_2\otimes \text{I}^{R_1}  \otimes \text{I}^{R_2}\rho^{S_1\ldots S_NR_1R_2} U_1^{\dagger}\otimes U_2^{\dagger}\otimes \text{I}^{R_1} \otimes \text{I}^{R_2})$ as $\rho^C$.
The reduced density matrix shared between $S_1 \ldots S_r$ and $R_1$ is given by
\begin{eqnarray}\label{Eq:Xi_onedef}
\xi^1_{j} &=& \text{tr}_{S_{r+1}\ldots S_NR_2}\Lambda_{S_1\ldots S_NR_1R_2}^C(\rho_{j,j'}^{S_1\ldots S_NR_1R_2})\nonumber \\
&=& (W_j^{S_1\ldots S_r}\otimes \text{I}^{R_1}) \text{tr}_{S_{r+1} \ldots S_NR_2}(\rho^C) \nonumber \\ && \hspace{10em}(W_j^{S_1\ldots S_r \dagger}\otimes \text{I}^{R_1})
\end{eqnarray}
where we have used the fact that any bipartite state, $\rho_{AB}$, satisfy  
\begin{eqnarray}
\text{tr}_A((U_A\otimes U_B) \rho_{AB} (U_A^{\dagger}\otimes U_B^{\dagger})) = U_B \text{tr}_A (\rho_{AB}) U_B^{\dagger}. \hspace{1em}
\end{eqnarray} 
The Hilbert-Schmidt decomposition of $\rho^1 = \text{tr}_{S_{r+1} \ldots S_NR_2}(\rho^C)$ on $ {\cal H}^{S_1\ldots S_rR_1} $ in the $S_1\ldots S_r:R_1$ bipartition  is given by 
\begin{eqnarray}\label{Eq:decompositionrho_one}
 \rho^1 = \frac{\text{I}^{S_1\ldots S_r}}{d_{S_1\ldots S_r}}\otimes \rho_{R_1}^1 + \sum_{k=0}^{d^2_{S_1\ldots S_r} - 1} r_k \mu_k^{S_1\ldots S_r}\otimes \text{I}^{R_1}\nonumber \\
 + \sum_{k=0}^{d^2_{S_1\ldots S_r} - 1} \sum_{l = 0}^{d^2_{R_1} - 1} t_{kl} \mu_k^{S_1\ldots S_r}\otimes \eta_l^{R_1},
\end{eqnarray}
where $\text{tr}_{S_1 \ldots S_r}\rho^1 = \rho_{R_1}^1$, $\mu_k$ and $\eta_l$ respectively are the generators of $SU(d_{S_1\ldots S_r})$ and $SU(d_{R_1})$, and where $\text{tr} \, \mu_k  = \text{tr} \,\eta_l= 0$ and $r_k$, $t_{kl}$ are real numbers. Using this form, the reduced density matrix of the output state is given by

\begin{eqnarray}
 \xi^1 = \frac{1}{d^2_{S_1\ldots S_r}}\sum_{j} \xi^1_{j} = \frac{\text{I}^{S_1\ldots S_r}}{d_{S_1\ldots S_r}}\otimes \rho_{R_1}^1,
\end{eqnarray}
where the second equality comes from the fact that $ \sum_jW_j\mu_k^{S_1\ldots S_r}W_j^{ \dagger} = d_{S_1\ldots S_r} (\text{tr}\mu_k^{S_1\ldots S_r}) \text{I} = 0 $. Since neither the CPTP map nor the unitary operators are acting on the part of the shared state in the receiver's side, $R_1$,  we have $\rho_{R_1}^1 = \rho^{R_1}$. Finally, we have
 \begin{equation}
  S(\xi^1) = \log d_{S_1\ldots S_r} + S(\rho^{R_1}),
 \end{equation}
 and similarly 
\begin{equation}
  S(\xi^2) = \log d_{S_{r+1}\ldots S_N} + S(\rho^{R_2}).
 \end{equation}

\noindent Note that in the case of arbitrary noise, the above equalities were inequalities as given in  (\ref{Eq:xi_inequality}). 

Let us now consider the third term in the R.H.S. of (\ref{Eq:maximize}). For example, if $x=1$, we have
 \begin{equation}
  \sum_j p_j S(\xi_j^1) = S(\rho^1),
 \end{equation}
where we use Eq. (\ref{Eq:Xi_onedef}) and the fact that unitary operations do not change the spectrum of any density matrix.

Interestingly, $S(\rho^1)$ does not depend on $W_j^{S_1 \ldots S_r}$ and $W_{j'}^{S_{r+1}\ldots S_N}$. It only depends on the fixed unitary operators $U_1^{S_1\ldots S_r}$ and the covariant channel, $\Lambda_{S_1\ldots S_NR_1R_2}^C$. The remaining task is to 
 minimize $S(\rho^1)$, by varying the $U_1^{S_1 \ldots S_r}$'s. Note that we have already shown that the first two terms in the R.H.S. of (\ref{Eq:maximize}) are independent of maximizations.
We now suppose that the minimum value of $S(\rho^1)$ is $S(\zeta^1)$ which will be achieved by setting $U^1_{min} = U_{min}^{S_1\ldots S_r}$. Similarly, for $x=2$, we have that the optimal $\sum_{j'} p_{j'}S(\xi^2_{j'})$ is $ S(\zeta^2)$, for the optimal unitary $U^2_{min} = U_{min}^{S_{r+1}\ldots S_N}$. Both the above inequalities can be  achieved by using orthogonal unitary operators applied with equal probabilities. We have therefore proved the following proposition.

\noindent{\bf Proposition 1:} \textit{For any covariant noisy channel between an arbitrary number of senders and  two receivers in a multiparty DC protocol, the capacity of LOCC-DC is bounded above by}
\begin{eqnarray}\label{Eq:Capacitydef_noise}
  \chi^{\text{LOCC}}_{noisy} = \log d_{S_1 \ldots S_N} + S(\rho^{R_1})  +S(\rho^{R_2})-\max_{x\in 1,2}S(\zeta^x), \nonumber \\ 
\end{eqnarray}
where $\zeta^x$ are given by
\begin{eqnarray}\label{Eq:zetaOneCovariant}
 \zeta^1 = \text{tr}_{S_{r+1}\ldots S_NR_2}\Lambda^C_{S_1\ldots S_NR_1R_2}(\rho^C_{min}),
 \end{eqnarray}
 and
 \begin{eqnarray}\label{Eq:zetaTwoCovariant}
 \zeta^2 = \text{tr}_{S_{1}\ldots S_rR_1}\Lambda^C_{S_1\ldots S_NR_1R_2}(\rho^C_{min}).
\end{eqnarray}
Here $\rho^C_{min} = \Lambda^C_{S_1\ldots S_N} (U^1_{min} \otimes U^2_{min}\otimes \text{I}^{R_1}  \otimes \text{I}^{R_2}\rho^{S_1\ldots S_NR_1R_2} U^{1 \dagger}_{min}\otimes U^{2 \dagger}_{min}\otimes \text{I}^{R_1} \otimes \text{I}^{R_2}) $

Depending on the specific covariant channels, the minimum unitaries, $U_{min}^1$ and $U_{min}^2$ can be obtained.  We find minimum unitaries for certain specific channels in the next section, where both covariant  as well as non-covariant channels will be considered. In Theorem 1, we proved that for an arbitrary noisy channel, the upper bound on the LOCC-DC capacity as given in inequality (\ref{Eq:maximize}) is further bounded above by the expression given in Eq. (\ref{Eq:Theorem_statement}). Proposition 1 shows that for covariant noisy channels, the two upper bounds are equal.

\section{Some examples of noisy quantum channels}\label{Sec:exmaple_noise}
In this section, we consider the shared state as the four-qubit GHZ state \cite{GHZ}, and consider different types of noisy channels.   
Undoubtedly, the GHZ state is one of the most important multiparty states, having maximal multiparty entanglement \cite{GM, GGM} as well as maximal violations of certain Bell inequalities \cite{GHZbellineq}. Moreover, it has  been successfully  realized in  laboratories by using several physical systems, including photons and ions \cite{GHZexp}. 
Our aim is to find the minimum unitary operators $U_{min}$ involved in $\zeta^1$ and $\zeta^2$ for different channels for this state, when the latter is used for LOCC-DC.

A four-qubit GHZ state shared between two senders, $S_1,S_2$ and two receivers, $R_1,R_2$ is given by
\begin{eqnarray}\label{Eq:GHZdef}
 &&|GHZ\rangle_{S_1S_2R_1R_2} = \frac{1}{\sqrt{2}}(|00\rangle_{S_1S_2}|00\rangle_{R_1R_2} \nonumber \\
&& \hspace{12em}+ |11\rangle_{S_1S_2}|11\rangle_{R_1R_2}).
\end{eqnarray}
We are now going to find out the  $U^{S_1}_{min}\otimes U^{S_2}_{min}$ that minimizes  $\max_{x\in 1,2}S(\zeta^x)$, where 
\begin{eqnarray}\label{Eq:zetaOne4}
 \zeta^1 = \text{tr}_{S_{2}R_2}\Lambda_{S_1S_2}(\tilde{\rho}^{S_1S_2R_1R_2}),
 \end{eqnarray}
 and
 \begin{eqnarray}\label{Eq:zetaTwo4}
 \zeta^2 = \text{tr}_{S_{1}R_1}\Lambda_{S_1S_2}(\tilde{\rho}^{S_1S_2R_1R_2}).
\end{eqnarray}
Here $\tilde{\rho}^{S_1S_2R_1R_2} = U^{S_1}_{min}\otimes U^{S_2}_{min}\otimes \text{I}^{R_1}\otimes \text{I}^{R_2} 
|GHZ\rangle\langle GHZ|_{S_1S_2R_1R_2}
U^{S_1 \dagger}_{min}\otimes U^{S_2 \dagger}_{min}\otimes \text{I}^{R_1}\otimes \text{I}^{R_2}$.
Note that $\Lambda_{S_1S_2}$ acts only on the senders' subsystems. We also denote $|GHZ\rangle\langle GHZ|$ as $\rho_{GHZ}$.

 To find the form of $U_{min}^{S_1}$ and  $U_{min}^{S_2}$, let us consider an arbitrary $2\times 2$ unitary matrix, acting on a sender's subsystem, given by
\begin{equation}\label{Eq:unitary_operator}
U^{S_i} =\left(
\begin{array}{cc}
a_ie^{i\theta_i^1} & \sqrt{1-a_i^2}e^{-i\theta_i^2} \\
-\sqrt{1-a_i^2}e^{i\theta_i^2} & a_ie^{-i\theta_i^1}
\end{array}\right),
\end{equation}
for $ i = 1,2$,
where $0\leq a_i\leq 1$ and $0 \leq \theta_i^1,  \theta_i^2 \leq \frac{\pi}{2} $. To find $\zeta^1$, we require only to manipulate the $U^{S_1}$, since $U^{S_2}$ is not involved in $\zeta^1$.  A similar statement is true for  $\zeta^2$.  

Let us now consider three classes of noisy channels, viz.
\begin{enumerate}
\item the amplitude damping,
\item phase damping, and
\item Pauli channels.
\end{enumerate}
Note that only the Pauli channel is a covariant one. In all the examples considered in this section, we consider that there are local channels which act on the individual channels running from the two senders to the two receivers. Note that from the perspective of the actual realizations, this is the more reasonable scenario.

These channels play important roles in the problem of decoherence \cite{Preskil}. The  amplitude damping channel has been used to model the spontaneous decay of a photon from an excited atomic state to its ground state,  while the phase damping one can correspond to scattering events. Pauli channels include a reasonably large class of quantum channels like the bit flip, and  depolarizing channels, and also play an important role in the problem of decoherence.

  
\subsection{Amplitude Damping Channel}\label{subsec:ampdamp}
A qubit in the  state $\rho$, after passing through the amplitude damping channel is given by
\begin{equation}
\rho \rightarrow {\cal A}_{\gamma}(\rho) = M_0\rho M_0^{\dagger} + M_1\rho M_1^{\dagger},
\end{equation} 
where the Kraus operators, $M_i$, $i= 0,1$ are given by
$$
M_0 = \left(
\begin{array}{cc}
1 & 0 \\
0 & \sqrt{1-\gamma}
\end{array}\right),
\,\,\,\,\,\,\,
M_1 = \left(
\begin{array}{cc}
0 & \sqrt{\gamma} \\
0 & 0
\end{array}\right),
$$
satisfying the condition
\begin{equation}
M_0^{\dagger}M_0 + M_1^{\dagger}M_1 = 1,
\end{equation}
with $0\leq\gamma\leq 1$.

In the dense coding scenario, the senders, $S_1$ and $S_2$, send their parts of the four-qubit GHZ state through local  amplitude damping channels, after encoding, and the corresponding output state is given by
\begin{eqnarray}\label{Eq:ampdampexpress}
\Lambda^{{\cal A}DC}(\rho_{GHZ}^{S_1S_2R_1R_2}) = \frac{1}{2}\{{\cal A}_{\gamma_1}(|0\rangle\langle 0|)\otimes {\cal A}_{\gamma_2}(|0\rangle\langle 0|)\nonumber \\ \otimes |00\rangle\langle 00| + {\cal A}_{\gamma_1}(|0\rangle\langle 1|)\otimes {\cal A}_{\gamma_2}(|0\rangle\langle 1|) \otimes |00\rangle\langle 11| \nonumber \\ + {\cal A}_{\gamma_1}(|1\rangle\langle 0|)\otimes {\cal A}_{\gamma_2}(|1\rangle\langle 0|) \otimes |11\rangle\langle 00|  \nonumber \\ +{\cal A}_{\gamma_1}(|1\rangle\langle 1|)\otimes {\cal A}_{\gamma_2}(|1\rangle\langle 1|) \otimes |11\rangle\langle 11|\}. \hspace{2em}
\end{eqnarray}
Here, \(\gamma_1\) and \(\gamma_2\) are the damping parameters for the two independent amplitude damping channels corresponding to the two channels from the senders to their corresponding receivers.  
Due to the symmetry of the GHZ state, it can be seen that $S(\zeta^2)$ takes the same functional form like $S(\zeta^1)$, when $\gamma_1$ and  $\gamma_2$ are interchanged.

By using the unitary operator given in Eq. (\ref{Eq:unitary_operator}), one can find that the eigenvalues of $\zeta^1$  are \begin{eqnarray}
\lambda_1 = \frac{1}{4}\left(1-\sqrt{f(a_1)}\right),\\
\lambda_2 = \frac{1}{4}\left(1+\sqrt{f(a_1)}\right),\\
\lambda_3 = \frac{1}{4}\left(1-\sqrt{g(a_1)}\right),\\
\lambda_4 = \frac{1}{4}\left(1+\sqrt{g(a_1)}\right),
\end{eqnarray} 
where $f(a) = 1-4\gamma_1(1-\gamma_1)a^4$ and $g(a) = 1-4\gamma_1(1-\gamma_1)(1-a^2)^2$. Note that the $\lambda_i$'s are independent of the $\theta_1^j$. 

\begin{figure}[h ]
 \includegraphics[width=0.9\columnwidth,keepaspectratio,angle=0]{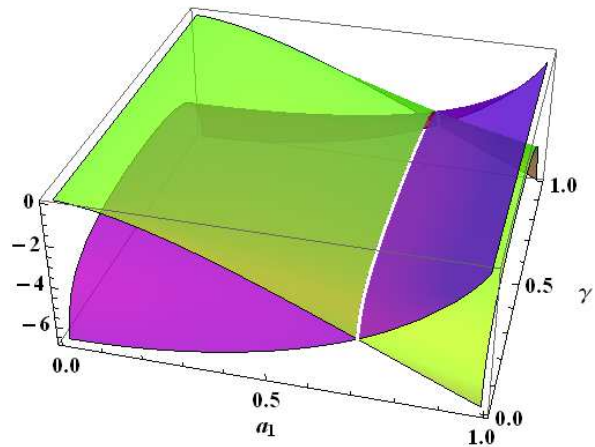}
\caption{(Color online.) Plots of the quantities $\frac{a_1^2}{\sqrt{f(a_1)}} \log \frac{1 - \sqrt{f(a_1)}}{1 + \sqrt{f(a_1)}} $ and  $\frac{1 - a_1^2}{\sqrt{g(a_1)}} \log \frac{1 - \sqrt{g(a_1)}}{1 + \sqrt{g(a_1)}}$, which are respectively the left-hand- and right-hand-sides of Eq. (\ref{Eq:extrcondition}), against \(a_1\) and \(\gamma\).  The green (gray in print) surface represents the first while  the  purple (dark in print) one is for the second expression.  The intersection line (white line)  is $a_1 = \frac{1}{\sqrt{2}}$, for all $\gamma$. The base axes are dimensionless, while the vertical axis is in bits.}
\label{fig:AmpDamp}
 \end{figure}

The minimization of $S(\zeta^1)  = -\sum_i \lambda_i\log \lambda_i \equiv F(a_1)$, say, is obtained by calculating 
\begin{equation}
\frac{dF(a_1)}{da_1} = 0,
\end{equation}
which lead to the relation given by
\begin{equation}\label{Eq:extrcondition}
\frac{a_1^2}{\sqrt{f(a_1)}} \log \frac{1 - \sqrt{f(a_1)}}{1 + \sqrt{f(a_1)}} = \frac{1 - a_1^2}{\sqrt{g(a_1)}} \log \frac{1 - \sqrt{g(a_1)}}{1 + \sqrt{g(a_1)}},
\end{equation}
Solutions of the above equation give the extrema.
In Fig. \ref{fig:AmpDamp}, we plot the L.H.S (left-hand-side, green surface) and R.H.S (purple surface) of Eq. (\ref{Eq:extrcondition}). The intersection line, $a_1 = \frac{1}{\sqrt{2}}$, of these two surfaces gives the solution of Eq. (\ref{Eq:extrcondition}).

To check whether it is minimum or not, we find 
\begin{eqnarray}
\left.\frac{d^2F(a_1)}{da_1^2}\right\vert_{a_1=\frac{1}{\sqrt{2}}} \hspace{-1.5em}= -\frac{\gamma(1-\gamma)}{\sqrt{(1-\gamma+\gamma^2)^3}}\bigg[\log\left(\frac{1-\sqrt{1-\gamma+\gamma^2}}{1-\sqrt{1+\gamma+\gamma^2}}\right)\nonumber \\\times(4 - 2\gamma(1-\gamma)) + 8\sqrt{1-\gamma +\gamma^2}\bigg],\hspace{2em}
\end{eqnarray}  
%
%
which is non-negative for all $\gamma$, at $a_1 = \frac{1}{\sqrt{2}}$, confirming the minimum. Therefore, the minimum of $S(\zeta^1)$ is obtained at  $a_1=\frac{1}{\sqrt{2}}$ and is given by
 $ 1 + H(\frac{1}{2}(1-\sqrt{1-\gamma_1 + \gamma_1^2}))$, where $H(x) = -x\log x - (1-x)\log(1-x)$ is the Shannon binary entropy of $x \in [0,1]$. Similarly, one can obtain the minimum of $S(\zeta^2)$. Note that there is a single extremal point obtained and the corresponding function is continuous, which implies that the local minimum obtained here is actually the global minimum.  
 Therefore, for the amplitude damping channel, if the input state is the $GHZ$ state, then the LOCC-DC capacity is given by
\begin{equation}
C_{{\cal A}DC}^{LOCC}  \leq 3 - \max_{x\in 1,2} H\left(\frac{1}{2}(1-\sqrt{1-\gamma_x + \gamma_x^2})\right).
\end{equation} 
Note that it is known that $C^{LOCC} = 3$, for the four-qubit GHZ state, with two receivers, in the case of noiseless channel \cite{DistriDC,Dense_multi},  and hence the capacity decreases in the presence of noise.

\subsection{Phase Damping Channel}\label{subsec:phasedamp}
In case of the phase damping channel, $\Lambda^{PD}$, the qubit in state $\rho$, changes as  
\begin{equation}
\Lambda^{PD}(\rho) = M_0\rho M_0^{\dagger} + M_1\rho M_1^{\dagger} + M_2\rho M_2^{\dagger},
\end{equation}
where the $M_i$'s are 
$$
M_0 = \left(
\begin{array}{cc}
\sqrt{1-p} & 0 \\
0 & \sqrt{1-p}
\end{array}\right),
$$
$$
M_1 = \left(
\begin{array}{cc}
\sqrt{p} & 0 \\
0 & 0
\end{array}\right),
\,\,\,\,\,\,\,
M_2 = \left(
\begin{array}{cc}
0 & 0 \\
0 & \sqrt{p}
\end{array}\right),
$$
with $0\leq p \leq 1$. Here we again assume that the noise is local on the senders' parts. 
In this case, the eigenvalues of $\zeta^1$, are given by
\begin{eqnarray}
\lambda_1 = \lambda_2 = \frac{1}{4}\left(1-\sqrt{f_P(a_1)}\right),\\
\lambda_3 = \lambda_4 = \frac{1}{4}\left(1+\sqrt{f_P(a_1)}\right),
\end{eqnarray}
where $f_P(a) = 1 - 4a^2(1-a^2)p(2-p)$. Like in the case of the amplitude damping channel, the minimization does not depend on the $\theta_i$'s. It is also clear from the concavity of the von Neumann entropy that maximizing $f_P(a_1)$ is enough to minimize $S(\zeta^1)$. Note that when $f_P(a_1)$ increases,  $\lambda_1$ and $ \lambda_2$ go close to zero while $\lambda_3$ and  $\lambda_4$ tend to $0.5$, which in turn minimize $S(\zeta^1)$. The second term in $f_P(a_1)$ is a positive quantity, the maximum value of $f_P(a_1)$ is $1$, when $a=0$ or $1$, and hence we have $S(\zeta^1) = 1$. Therefore, for the phase damping channel, we get
\begin{equation}
C^{\text{LOCC}}_{FP} \leq 3,
\end{equation}
which is independent of the parameters of the channel.

\subsection{Pauli Noise: A Covariant Channel}\label{Paulinoise}
Pauli noise is an example of a covariant noise, which satisfies the covariant condition, given in Eq. (\ref{Eq:Covariant_def}). When an arbitrary qubit state, is passed through the channel with Pauli noise  \cite{covariancenoise,PauliChannelgen}, the state is transformed as
\begin{equation}
\label{Eq:Paulinoise}
 \Lambda^P(\rho)=\sum_{m,n=0}^{1} q_{mn} W_{mn} \rho W_{mn}^{\dagger}
\end{equation}
where
 $\{W_{mn}\}$ are the well-known Pauli spin matrices and the identity operator, i.e,
$$
W_{01} = \sigma_1=\left(
\begin{array}{cc}
0 & 1 \\
1 & 0
\end{array}\right)
,\,\,\,
W_{11} = \sigma_2=\left(
\begin{array}{cc}
0 & -i \\
i & 0
\end{array}\right)
,$$
$$
W_{10} = \sigma_3=\left(
\begin{array}{cc}
1 & 0 \\
0 & -1
\end{array}\right)
,\,\,\,
W_{00} = \sigma_0=\left(
\begin{array}{cc}
1 & 0 \\
0 & 1
\end{array}\right)
.$$
Consider a four-qubit state, $\rho^{S_1S_2R_1R_2}$, shared between two senders and two receivers. After passing through the Pauli channel, it transforms as
\begin{eqnarray}\label{Eq:pauli_def}
& & \Lambda^P_{S_1S_2R_1R_2}(\rho^{S_1S_2R_1R_2}) = \sum_{mn} q_{mn} \nonumber \\
& & \left(\sigma_{m}^{S_1}\otimes\sigma_{n}^{S_2}\otimes \text{I}^{R_1} \otimes \text{I}^{R_2}\right) \rho^{S_1S_2R_1R_2} \left(\sigma_{m}^{S_1}\otimes\sigma_{n}^{S_2}\otimes\right. \nonumber \\
& & \left. \text{I}^{R_1} \otimes \text{I}^{R_2}\right),
\end{eqnarray}
where $\sum_{mn}q_{mn} = 1$. Depending on the choice of $q_{mn}$, the channel can be correlated or uncorrelated. We  deal with the fully-correlated Pauli channel, i.e, when  $q_{mn} = q_m \delta_{mn}$. Eq. (\ref{Eq:pauli_def}) in this case reduces to
\begin{eqnarray}\label{Eq:fully_Pauli_def}
& & \Lambda^{fP}_{S_1S_2R_1R_2}(\rho^{S_1S_2R_1R_2}) = \sum_{m} q_{m} \nonumber \\
& & \left(\sigma_{m}^{S_1}\otimes\sigma_{m}^{S_2}\otimes \text{I}^{R_1} \otimes \text{I}^{R_2}\right) \rho^{S_1S_2R_1R_2} \left(\sigma_{m}^{S_1}\otimes\sigma_{m}^{S_2}\otimes \right. \nonumber \\
& &\left. \text{I}^{R_1} \otimes \text{I}^{R_2}\right).
\end{eqnarray}
Let us find out the $U_{min}$ for the four-qubit GHZ state shared between two senders and two receivers, in the presence of the fully-correlated Pauli noise as in Eq. (\ref{Eq:fully_Pauli_def}).
From the symmetry of the GHZ state, we have  $S(\zeta^1) = S(\zeta^2)$. The eigenvalues of $\zeta^1$ are given by
\begin{equation}
 \lambda_1 = \lambda_2 = \frac{1}{4}\left(1 - \sqrt{g(a_1,\theta_1^1,\theta_1^2)}\right),
\end{equation}
\begin{equation}
 \lambda_3 = \lambda_4 = \frac{1}{4}\left(1 + \sqrt{g(a_1,\theta_1^1,\theta_1^2)}\right),
\end{equation}
where
\begin{eqnarray}\label{Eq:gtheta}
 \tilde{g}(a,\theta)&\equiv& g(a,\theta_1, \theta_2) = (q_0 - q_1 - q_2 + q_3)^2  \nonumber \\
 & &  \hspace{-2em} + f_1(a)[8q_1q_2 + 8q_0q_3 - 4(q_0 + q_3)(q_1+q_2)\nonumber \\
 & & \hspace{4em} - 4(q_1 - q_2)(q_0 - q_3)\cos(2(\theta_1+\theta_2))]\nonumber \\
\end{eqnarray}
and $f_1(a) = 2a^2(-1 + a^2)$. Arguing in the same way as in other cases, it is enough to maximize  $\tilde{g}(a,\theta)$, with $\theta = \theta_1 + \theta_2$, in order to minimize $S(\zeta^1)$. 
To find the extremum of $\tilde{g}(a,\theta)$, we have to solve 
\begin{equation}
 \frac{\partial \tilde{g}(a,\theta)}{\partial a} = 0,
\end{equation}
and
\begin{equation}
 \frac{\partial \tilde{g}(a,\theta)}{\partial \theta} = 0,
\end{equation}
which give the extremum value at $a = a_0 \equiv 0 \, \text{or} \, \frac{1}{\sqrt{2}}$, and $\theta =\theta_0 \equiv \frac{n \pi}{2}$, where $n \in \mathbb{Z}$.  $\tilde{g}(a,\theta)$ is a function of the noise parameters $\{q_m\}$, and to find the extremum, without loss of generality, we assume an ordering of those parameters, i.e.,
we assume
\begin{equation}\label{Eq:ordering}
 q_0 \geq q_2 \geq q_1 \geq q_3.
\end{equation}
And $\tilde{g}(a,\theta)$ is maximum, when  
\begin{equation}
\left.\frac{\partial^2\tilde{g}(a,\theta)}{\partial a^2}\right|_{a_0,\theta_0},\,\, \left.\frac{\partial^2\tilde{g}(a,\theta)}{\partial {\theta}^2}\right|_{a_0,\theta_0} < 0,
\end{equation}
\begin{equation}
\left.\left(\frac{\partial^2\tilde{g}}{\partial a \partial \theta}\right)^2\right|_{a_0,\theta_0} < \frac{\partial^2\tilde{g}}{\partial a^2}\left.\frac{\partial^2\tilde{g}}{\partial {\theta}^2}\right|_{a_0,\theta_0},
\end{equation}
 are satisfied simultaneously. For the above  choice of $q_m$, the maximum value of $\sqrt{\tilde{g}(a,\theta)}$ is $|q_0 -q_1 + q_2- q_3|$, which will be achieved,  when $a = \frac{1}{\sqrt{2}}$ and $\theta $ is odd multiple of $\frac{\pi}{2}$, $S(\zeta^1) = H(q_0 + q_2) + 1$,  and $U_{min}^{S_1}$ is given by
$$
U_{min}^{S_1} =\frac{1}{\sqrt{2}}\left(
\begin{array}{cc}
e^{i\theta_1^1} & -ie^{i\theta_1^1} \\
-ie^{-i\theta_1^1} & e^{-i\theta_1^1}
\end{array}\right).
$$
If we take another ordering of $\{q_m\}$, for e.g., $q_1\geq q_2 \geq q_0 \geq q_3$, we have $S(\zeta^1) = H(q_1 + q_2) + 1$, and the unitary operator, in this case, is given by
$$
U_{min}^{S_1} =\left(
\begin{array}{cc}
0 & e^{i\theta_1^1} \\
e^{-i\theta_1^1} & 0
\end{array}\right).
$$
The above two cases indicate that the minimum entropy depends on the ordering of $q_m$, involved in the channel with Pauli noise. In general, when the shared state is the GHZ state, the capacity is bounded above by $3 - H(b_1+b_2)$, where $\{b_m\}_{m=1}^4$, is an  arrangement of $\{q_m\}$ in descending order.

Instead of fully correlated Pauli noise, if we now assume that the $q_{mn}$ is arbitrary, the strategy of fully correlated Pauli noise can also be applied in this case. Suppose, $p_m = \sum_n q_{mn}$ and $r_n = \sum_m q_{mn}$. Then the capacity is bounded above as
\begin{equation}
C_{Pauli}^{LOCC} \leq 3 - \max \{H(b_1+b_2), H(c_1 + c_2)\},
\end{equation}
where $\{b_m\}_{m=1}^4$ and $\{c_n\}_{n=1}^4$ are the sets $\{p_m\}$ and $\{r_n\}$ in descending order.

\section{Generalized Geometric Measure}\label{Sec:GGMDef}
\label{subsec:GGM}

We now define a genuine multipartite entanglement measure called the generalized geometric measure \cite{GGM} (cf. \cite{GM}). An $N$-party pure state is said to be genuinely multiparty entangled if it is non-separable under all bipartitions. For such states, one can define a multipartite entanglement measure based on the distance from the set of all multiparty states that are not genuinely multiparty entangled.

The GGM of  an \(N\)-party pure quantum state, \(|\phi_N\rangle\), is defined as \vspace{-2mm}
\begin{equation}
{\cal E} ( |\phi_N\rangle ) = 1 - \Lambda^2_{\max} (|\phi_N\rangle ),
\end{equation}
where  \(\Lambda_{\max} (|\phi_N\rangle ) =
\max | \langle \chi|\phi_N\rangle |\), with  the maximization being over all pure states \(|\chi\rangle\)
that are not genuinely \(N\)-party entangled.
It reduces to \cite{GGM} 
\begin{equation}\label{Eq:GGMdef}
{\cal E} (|\phi_N \rangle ) =  1 - \max \{\lambda^2_{{\cal A}: {\cal B}} |  {\cal A} \cup  {\cal B} =
\{1,2,\ldots, N\},  {\cal A} \cap  {\cal B} = \emptyset\},
\end{equation}
where \(\lambda_{{\cal A}:{\cal B}}\) is  the maximal Schmidt coefficient in the \({\cal A}: {\cal B}\)
bipartite split  of \(|\phi_N \rangle\).

\section{Multipartite entanglement and dense coding for more than one receiver}\label{Sec:Relationship_GGM_Capacity}
In this section, we establish a relation between the  capacities of LOCC-DC of four-qubit pure states with two senders and two receivers  and their generalized geometric measure $({\cal E})$. 
 Specifically, we will estimate the ordering of the GGMs between the generalized GHZ state and an arbitrary four-qubit pure state, when both of them have equal LOCC dense coding capacities.

Note that although the exact capacity of dense coding by LOCC for arbitrary multiparty pure state is not known, it was shown \cite{DistriDC,Dense_multi} that the exact  capacity is $3$ for the four-qubit GHZ state, given by $|GHZ\rangle = \frac{1}{\sqrt{2}}(|0000\rangle + |1111\rangle)$. In case of the gGHZ state, which is given by $|gGHZ\rangle = \alpha |0000\rangle + \sqrt{1-\alpha^2}e^{i\phi}|1111\rangle$, the capacity of LOCC-DC is bounded above by
\begin{equation}
{\cal B}^{LOCC}(|gGHZ\rangle) = 2+ H(\alpha).
\end{equation}  

From the intuition obtained from bipartite non-maximally entangled states, we conjecture here that the capacity of LOCC-DC for the gGHZ state saturate the upper bound, ${\cal B}^{LOCC}$. With this assumption,  we have the following theorem.   

\noindent{\bf Theorem 2:} \textit{ Consider a multiparty DC protocol where there are two senders and two receivers, and where the channels from the senders to the receivers are noiseless. In this case if a four-qubit gGHZ state and an arbitrary four-qubit pure  state have equal capacities of LOCC-DC, then the  gGHZ state possesses  less genuine multiparty entanglement than that of the arbitrary state i.e, we have}
\begin{equation}
  {\cal E}(|\psi \rangle) \geq {\cal E}(|gGHZ\rangle),
 \end{equation}
\emph{if (i) $S(\rho^{R_1}) \leq S(\rho^{S_1R_1})$, i.e., the reduced state, $\rho^{S_1R_1}$, has more disorder than its local subsystem, $\rho^{R_1}$, and (ii) the maximum eigenvalue required for GGM is obtained from the density matrix, $\rho^{R_2}$. Similar conditions can be obtained by interchanging $S_1$ and $R_1$ with $S_2$ and $R_2$ respectively.}
 
 \texttt{Proof:}
 As argued above, it is plausible that for the $gGHZ$ state,
\begin{equation}
\label{Eq:Capacity_def_gGHZ}
C^{LOCC}_{gGHZ} = 2+ H(\alpha).
\end{equation}
For an arbitrary four-qubit pure state, $|\psi\rangle$, shared between the senders $S_1,S_2$ and receivers $R_1,R_2$, the upper bound of the capacity of LOCC-DC is given by
\begin{eqnarray}
C^{LOCC}_{\psi} \leq {\cal B}^{LOCC}(|\psi\rangle) = 2+ S(\rho^{R_1}) + S(\rho^{R_2})\nonumber \\  - S(\rho^{S_1R_1}),
\end{eqnarray}
where $S(\rho^{R_i}), i = 1,2$, and $S(\rho^{S_1R_1})$ are the reduced density matrices of $|\psi\rangle$.

Note that for pure state $S(\rho^{S_1R_1}) = S(\rho^{S_2R_2})$. Let us now assume that the LOCC-DC capacities for $|\psi\rangle$ and the gGHZ state are equal, so that 
\begin{eqnarray}
C_{gGHZ}^{LOCC} &=& 2+ H(\alpha) = C^{LOCC}_{\psi}\nonumber \\  &\leq& 2+ S(\rho^{R_1}) + S(\rho^{R_2}) - S(\rho^{S_1R_1}),
\end{eqnarray} 
which implies $H(\alpha) \leq S(\rho^{R_2})$, provided $S(\rho^{R_1}) \leq S(\rho^{S_1R_1})$.

This implies that
\begin{equation}\label{Eq:Condition_GGM_prof}
 \alpha \geq \lambda^{R_2},
\end{equation} 
where $\lambda^{R_2}$ is the maximum eigenvalue of $\rho^{R_2}$.

The GGMs of the gGHZ and the arbitrary four-qubit pure state are respectively given by 
\begin{equation}
 {\cal E}(|gGHZ\rangle) = 1 - \alpha,
\end{equation}
\begin{equation}
 {\cal E}(|\psi\rangle) = 1 - \lambda^{R_2},
\end{equation}
provided that $\lambda^{R_2}$ is the maximum eigenvalue among all the eigenvalues of its single site and two site density matrices. Then, by using (\ref{Eq:Condition_GGM_prof}), we get 
\begin{equation*}
{\cal E}(|\psi\rangle) \geq {\cal E}(|gGHZ\rangle).
\end{equation*}
Hence the proof. \hfill $\blacksquare$


While the above theorem has been stated for two senders and two receivers, simple changes in the premises render it valid for the case of multiple senders and two receivers.

One should stress here that if the DC protocol involves several senders and a single receiver, it has recently been shown that the gGHZ state requires to be more multiparty entangled than an arbitrary four-qubit state if they both want to have equal DC capacities in a noiseless scenario \cite{Noiseinvert}.
 Here we show that changing the number of receivers from one to two can alter the hierarchy with respect to the multiparty entanglement and the multiparty DC capacity among four-qubit states and the gGHZ state.

\begin{figure}[t]
 \includegraphics[width=0.7\columnwidth,keepaspectratio,angle=270]{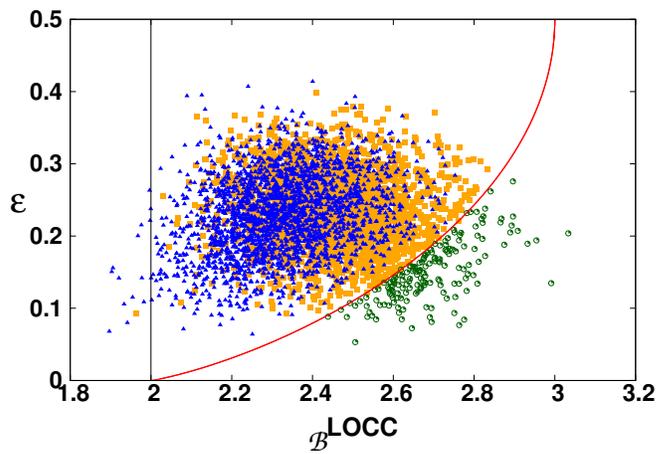}
\caption{(Color online.) Noiseless case: How does a general four-qubit pure state compare with the gGHZ states? We randomly generate $5\times 10^4 $ four-qubit pure states uniformly with respect to the corresponding Haar measure, and their GGM is plotted as the abscissa while ${\cal B}^{\text{LOCC}}$ is plotted as the ordinate. The red solid line represents the gGHZ states. Among the states generated randomly, $47.6\%$ (blue triangles) satisfy both the conditions in Theorem 2, $49\%$ (orange squares) violate either of the conditions, but still falls above the gGHZ line. Green circles represent $3.4\%$ states which violate the conclusion of Theorem 2. The line at abscissa equals to 2 corresponds to the capacity achievable without prior shared entanglement. The vertical axis is dimensionless, while the horizontal one is in bits. }
\label{fig:GGM_Blocc}
 \end{figure}

To visualize the above theorem (Theorem 2), and to check the relevance of the imposed conditions,  we randomly generate $10^5$ arbitrary four-qubit pure states, Haar-uniformly on that space.
In Fig. \ref{fig:GGM_Blocc},  the GGM (${\cal E}$) is plotted against the upper bound,  ${\cal B}^{\text{LOCC}}$, of the LOCC-DC capacity for the generated states. The red line represents the gGHZ states. Among the randomly generated states, $47.6\%$ states  (blue triangles) satisfy both the conditions $(i)$ and $(ii)$ of Theorem 2. Interestingly however, $49\%$ states (orange squares) violate at least one of the above conditions, and yet reside above the gGHZ line i.e., satisfy the conclusion of Theorem 2. And only $3.4\%$ of the total violate the conclusion of Theorem 2 (green circles).

\subsection{The Noisy Case}
\label{SubSec:Relationship_GGM_Capacity_noisy}

\begin{figure}
\begin{center}
 \includegraphics[width=0.675\columnwidth,keepaspectratio,angle=270]{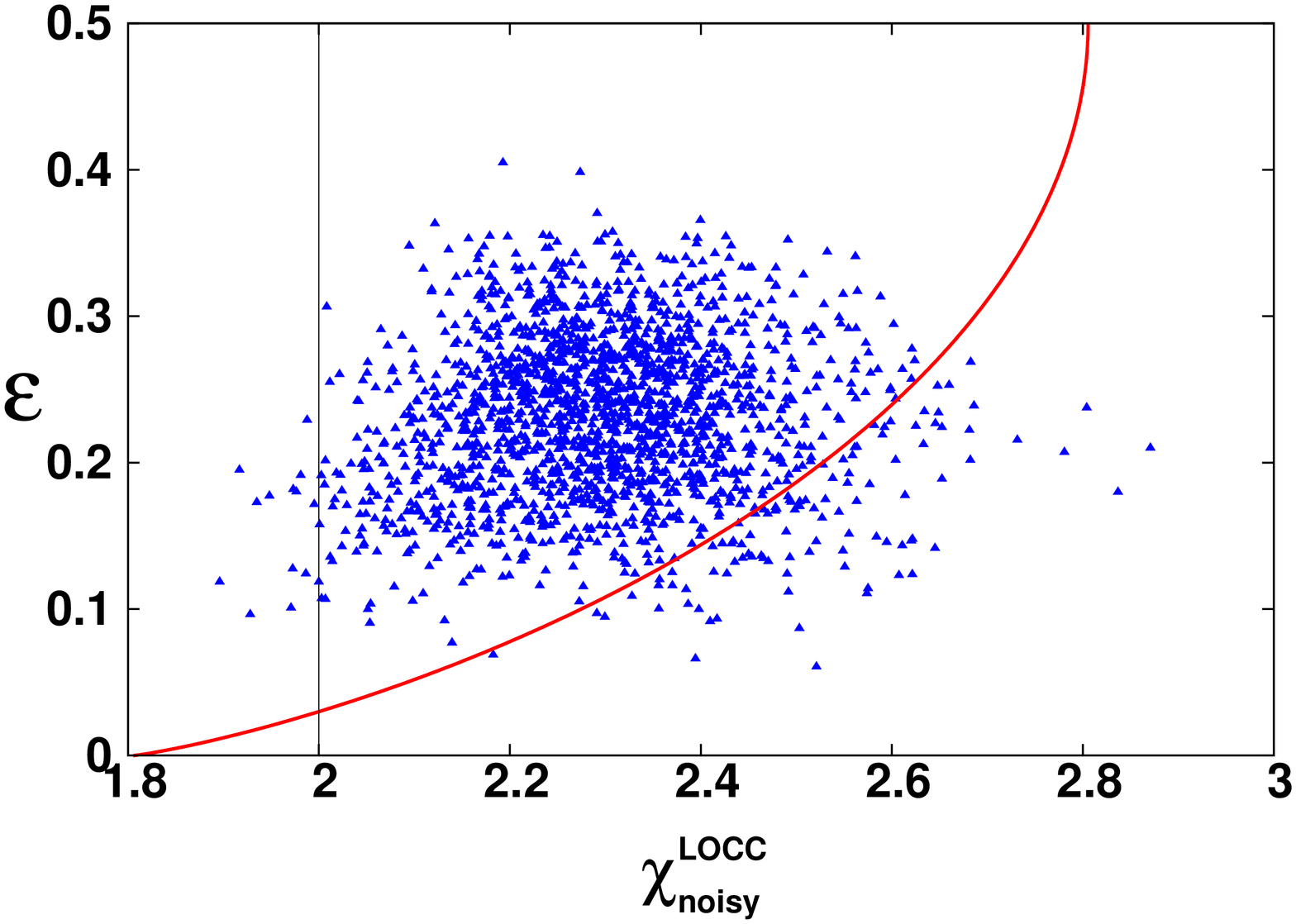}
 \includegraphics[width=0.675\columnwidth,keepaspectratio,angle=270]{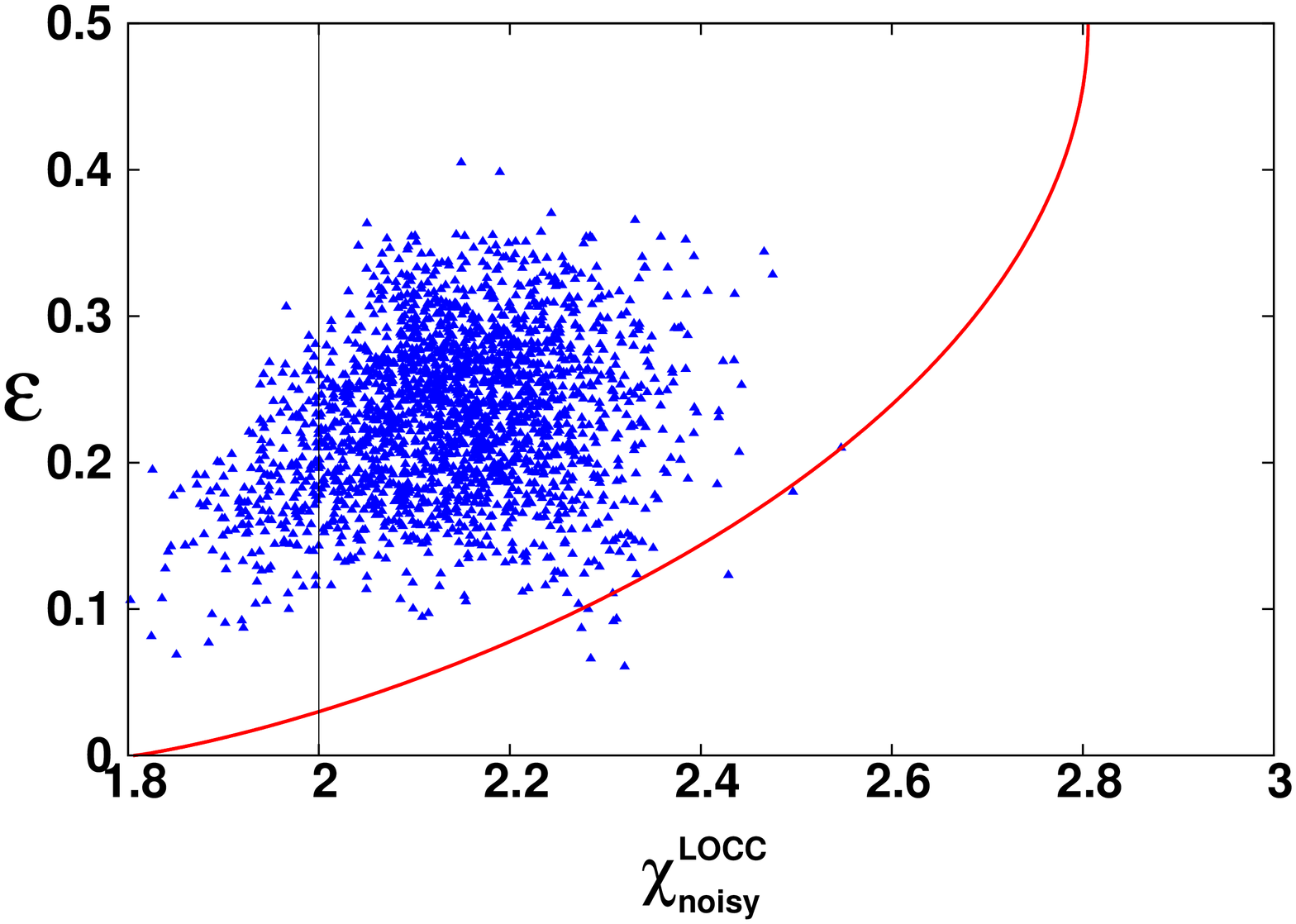}
 \caption{(Color online.) Fully correlated Pauli noise: The gGHZ states are again better than a significant fraction of states. We plot the GGM as the ordinate and  $\chi^{LOCC}_{noisy}$ as the abscissa for $5\times 10^4$ randomly generated four-qubit pure states uniformly with respect to the corresponding Haar measure for low (top panel) and high (bottom panel) full correlated Pauli noise. In the top panel, $q_0 = 0.93, q_1 = 0.01, q_2 = 0.02, q_3 = 0.04$, while in the bottom panel, we choose $q_0 = 0.485, q_1 = 0.015, q_2 = 0.015,  q_3 = 0.485$. In the presence of high noise, almost all states are bounded by the four-qubit gGHZ states (red solid line). A significant fraction of the generated states lie above the gGHZ line even for low noise. It indicates that the gGHZ state is more robust against noise as compared to an arbitrary four-qubit pure state. The lines at abscissa equals to 2 correspond to the capacity achievable without prior shared entanglement. The vertical axis is dimensionless, while the horizontal one is in bits.}
\label{fig:GGM_Chi_LOCC_noise} 
 \end{center}
 \end{figure}

We now try to find  a relation between the GGM and the maximal classical information transfer by LOCC, as quantified by $\chi_{noisy}^{LOCC}$ given in Eq. (\ref{Eq:Capacitydef_noise}), under fully correlated Pauli noisy channel. We randomly generate $5\times 10^4$  four-qubit pure states Haar-uniformly on the state space, and calculate the $\chi_{noisy}^{LOCC}$, for the states under Pauli noise. We do the same for the generalized GHZ states. 
We  choose two  sets of noise parameters: $(i)$ parameters that lead to a state which is close to the state of the noiseless case, and we refer it  as the low noise case, and $(ii)$ parameters which take the state close to the maximally mixed state, and we refer to it as the high noise case. Our aim is to connect the LOCC-DC capacity in the presence of Pauli noise,  and  multiparty entanglement, as quantified by the GGM, of the initially shared state.
For the low noise case, we choose the noise parameters as  $q_0 = 0.93,\, q_1 = 0.01,\, q_2 = 0.02$, and $q_3 = 0.04$, and plot the GGM against $\chi_{noisy}^{LOCC}$. 
For the high noise case, we choose $q_0 = 0.485,\, q_1 = 0.015,\, q_2 = 0.015,\, q_3 = 0.485$. 
The plots are presented in  Fig. \ref{fig:GGM_Chi_LOCC_noise}.  In the high-noise case, the upper bound on the LOCC-DC capacity, as expected,  suggests that most of the states have capacities  which are lower than the capacity achieved by the classical protocol.  In the noiseless as well as the low noise scenarios, we see that there exists a set of  states  which is not bounded by the gGHZ line, while such states are almost absent in the presence of higher amounts of noise (see Fig. \ref{fig:GGM_Chi_LOCC_noise}).  It suggests that the gGHZ state is more robust to noise 
among  four-qubit pure states.



\ 


For the case of multiple senders and a single receiver, the gGHZ state
changes its role as one increases noise in the channel that carries the
encoded quantum systems from the senders to the receiver \cite{Noiseinvert}. Precisely,
the gGHZ state requires less multiparty entanglement (as quantified by
GGM) than a generic state to be equal in dense coding capacity with the
generic state, if the channels are noisy. The opposite is true when the
channels are noisy. Here we see that if there are two receivers in the
protocol, there is no such role reversal. The gGHZ state requires less
multiparty entanglement than a generic state to have the same LOCC dense
coding capacity as the generic state. Note that this statement in under
the assumption that the upper bounds on the LOCC-DC capacities faithfully
mirror the qualitative features of the actual capacities.

\section{Conclusion}\label{Sec:Conclusion}
The dense coding protocol is a quantum communication scheme which demonstrates that the classical information  can be transferred via quantum states more efficiently than any classical protocol. The ``Holevo bound'' is applied to obtain the capacities, when there is a single sender and a single receiver as well as when there are multiple senders and a single receiver. Capacities are known for both noiseless and noisy channels. However, realistic scenarios of a communication protocol should involve multiple senders and multiple receivers. The difficulty in such generalization is due to the nonexistence, hitherto, of a  Holevo-like bound in the multipartite decoding process in the many-receivers scenario in the case of noisy channels. In this paper, we address the problem of estimating the dense coding capacity, when there are an arbitrary number of senders and two receivers. In particular, we find an upper bound on the classical capacity of the multipartite quantum channel, when the senders and receivers share a multiparty quantum state and noisy channels, and the receivers are allowed to perform only local quantum operations and classical communication. 
A compact form of the upper bound on the capacity is obtained when the noisy channels are covariant. 
When the four-party shared state is the GHZ state, several paradigmatic noisy channels are considered and the upper bounds on the capacities are determined. Finally, we connect the capacity of dense coding with a multiparty entanglement measure, both in the noiseless and noisy scenarios.


\begin{thebibliography}{99}
\bibitem{Female} In this paper, the senders are female, and receivers are male.
\end{thebibliography}
\end{document}